\documentclass{v15cosmo}
\def\be{\begin{equation}}
\def\ee{\end{equation}}
\def\bea{\begin{eqnarray}}
\def\eea{\end{eqnarray}}

\usepackage{wrapfig}

\usepackage{color}
\usepackage{hyperref} 
\hypersetup{colorlinks=true,citecolor=blue}

\def \bea {\begin{eqnarray}}
\def \ena {\end{eqnarray}}

\def	\B		{\,{\rm B}}

\def    \exp 		{\,{\rm exp}}

\def	\K		{{\,\rm K}}

\def	\sp		{\,{\rm sp}}
\def	\ed		{\,{\rm ed}}

\def	\H		{{\rm H}}

\def	\ba		{{\bf a}}
\def	\be		{{\bf e}}
\def	\bJ		{{\bf J}}

\font\mib=cmmib10

\def \bmu {{\hbox {\mib\char"16}}}
\def \bomega {\hbox {\mib\char"21}}

\newcommand{\Photo}{}
\begin{document}

\vspace*{2cm}
\title{Anomalous Microwave Emission from Spinning Dust and its Polarization Spectrum}
\author{Thiem Hoang}
\address{Institute of Theoretical Physics, Goethe  Universit$\ddot{\rm a}$t Frankfurt, Germany \\
Current address: Canadian Institute for Theoretical Astrophysics, University of Toronto, Canada}

\maketitle\abstracts{
Nearly twenty years after the discovery of anomalous microwave emission (AME) that contaminates to the cosmic microwave background (CMB) radiation, its origin remains inconclusive. Observational results from numerous experiments have revealed that AME is most consistent with spinning dust emission from rapidly spinning ultrasmall interstellar grains. In this paper, I will first review our improved model of spinning dust, which treats realistic dynamics of wobbling non-spherical grains, impulsive interactions of grains with ions in the ambient plasma, and some other important effects. I will then discuss recent progress in quantifying the polarization of spinning dust emission from polycyclic aromatic hydrocarbons. I will finish with a brief discussion on remaining issues about the origins of AME.}

\section{Introduction}
Anomalous microwave emission (AME) was discovered about twenty years ago during analysis of data from the cosmic background explorer (COBE) satellite \cite{Kogut:1996p5303}. Its detections were quickly reported in the data sets from Saskatoon \cite{1997ApJ...482L..17D}, OVRO \cite{Leitch:1997p7359}, the 19~GHz survey \cite{deOliveiraCosta:1998p4707}, and Tenerife \cite{deOliveiraCosta:1999p4706}. Initially, AME was identified as thermal bremsstrahlung from ionized gas correlated with dust \cite{Kogut:1996p5303} and presumably produced by photoionized cloud rims \cite{1999ASPC..181..253M}. This idea was scrutinized in Draine \& Lazarian~\cite{1998ApJ...508..157D} and criticized on energetic grounds. Poor correlation of H$\alpha$ with 100~$\mu$m emission also argued against the free-free explanation \cite{1999ASPC..181..253M}. Later, de-Oliveira Costa \cite{deOliveiraCosta:2000p4667} used Wisconsin H-Alpha Mapper (WHAM) survey data and established that the free-free emission ``is about an order of magnitude below Foreground X over the entire range of frequencies and latitudes where it is detected''. The authors concluded that the Foreground X cannot be explained as the free-free emission. Additional evidence supporting this conclusion has come from a study at 5, 8 and 10~GHz by Finkbeiner et al.~\cite{2002ApJ...566..898F} of several dark clouds and H II regions, two of which show a significantly rising spectrum from 5 to 10 GHz. 

For the last several years, we have witnessed significant progress in observational studies of AME, aiming to better characterize Galactic foregrounds for precision cosmology. To date, spinning dust emission has been detected in a wide range of astrophysical environments, including general ISM \cite{Gold:2009p5268,Gold:2011p5261,PlanckCollaboration:2011p515}, star forming regions in the nearby galaxy NGC 6946 \cite{Scaife:2010p569,2012ApJ...754...94T}, and Perseus and Ophiuchus clouds \cite{2008MNRAS.391.1075C,2011MNRAS.418.1889T}. Planck results have been interpreted as showing a microwave emission excess from spinning dust in the Magellanic Clouds \cite{2010A&A...523A..20B,PlanckCollaboration:2011p515}. On the theoretical and modeling front, the original model of spinning dust \cite{1998ApJ...508..157D} has been improved substantially to incorporate important physical effects disregarded in the original model. In this short paper, I will review our comprehensive model of spinning dust, which treats realistic rotational dynamics of wobbling grains, and recent progress in quantifying the polarization of spinning dust emission.
 
\section{Models of Spinning Dust Emission}
\subsection{Original Model: Draine and Lazarian}

Below are the essential assumptions in Draine and Lazarian (DL98)\cite{1998ApJ...508..157D} model:

(i) The smallest PAH particles of a few \AA~ are expected to be planar. The grain size $a$ is defined as the radius of an equivalent sphere of the same mass. PAHs are assumed to be planar, disk-like for $a<a_{2}$ and spherical for $a \ge a_{2}$. The value of $a_{2}=6$\AA~ is adopted.

(ii) PAHs usually have electric dipole moment $\bmu$ arising from asymmetric polar molecules or substructures ({\it intrinsic dipole moment}) and from the asymmetric distribution of grain charge. The latter is shown to be less important.

(iii) The grain spins around its symmetry axis $\hat{\ba}_{1}$ with angular momentum $\bJ$ parallel to $\hat{\ba}_{1}$, and $\bJ$ is isotropically oriented in space.

(iv) For a fixed angular momentum, the spinning grain emits electric dipole radiation at a {\it unique} frequency mode $\nu$, which is equal to the rotational frequency, i.e., $\nu=\omega/2\pi$.

(v) A grain in the gas experiences collisions with neutral atoms and ions, interacts with passing ions (plasma-grain interactions), emits infrared photons following UV absorption, and emits rotational radiation due to rotating electric dipole. All these processes result in the damping and excitation of grain rotation, i.e., they change grain angular momentum $J$ and velocity $\omega$.

(vi) Due to the excitation of various aforementioned processes, the grain angular velocity randomly fluctuates and its distribution can be approximated as the Maxwellian distribution function $f_{\rm Mw}(\omega)$.

(vi) The total emissivity per H atom of the electric dipole radiation from spinning dust at the frequency $\nu$ is given by
\bea
\frac{j_{\nu}}{n_\H} =\frac{1}{4\pi}\frac{1}{n_\H}
\int_{a_{\min}}^{a_{\max}} da {dn\over da} 
4\pi \omega^2 f_{\rm Mw}(\omega) 2\pi 
\left(\frac{2\mu_{\perp}^{2}\omega^{4}}{3c^{3}}\right)~~~,
\label{eq:504}
\ena
where $n_{\H}$ is the density of H nuclei, $\mu_{\perp}$ is the electric dipole moment perpendicular to the rotation axis, and $dn/da$ is the grain size distribution function with $a_{\min}=0.35$\AA~ to $a_{\max}=100$\AA.

\begin{wrapfigure}{r}{0.45\textwidth}
\vspace*{-2.0cm}
\includegraphics[width=0.45\textwidth]{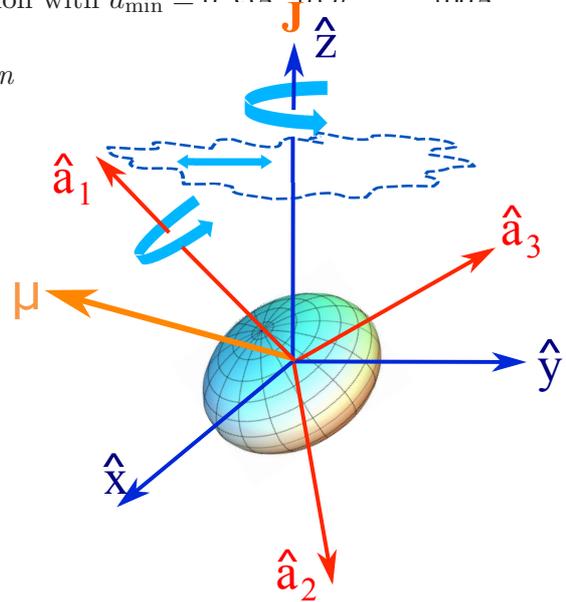}
\caption{\small Rotational configuration of a triaxial ellipsoid characterized by three principal axes $\hat{\ba}_{1}\hat{\ba}_{2}\hat{\ba}_{3}$ in the inertial frame $\hat{\bf x}\hat{\bf y}\hat{\bf z}$. Grain angular momentum $\bJ$ is conserved in the absence of external torques and directed along $\hat{\bf z}$. The torque-free motion of the grain comprises the rotation around $\hat{\ba}_{1}$, the precession of $\hat{\ba}_{1}$ around $\bJ$, and the wobbling motion of $\hat{\ba}_{1}$ relative to $\bJ$. The dipole moment $\bmu$ is fixed to the grain body.}
\label{fig:3Dgrain}
\end{wrapfigure}
\subsection{Improved Model: Hoang, Draine, and Lazarian}
Hoang et al.~\cite{Hoang:2010jy} improved the DL98 model by accounting for a number of fundamental physical effects. The main improvements in this model of spinning dust emission are summarized as follows.

(a) Disk-like grains rotate with their symmetry axis $\hat{\ba}_{1}$ not perfectly aligned with angular momentum $\bJ$. The disaligned rotation of $\bJ$ with $\hat{\ba}_{1}$ causes the wobbling motion of the grain principal axes with respect to $\bJ$ due to internal thermal fluctuations.

(b) The power spectrum of multiple emission frequency modes of a freely spinning grain is computed using the Fourier transform.

(c) The distribution function of grain angular momentum $J$ and velocity $\omega$  are obtained exactly using the Langevin equation (LE) for the evolution of $\bJ$ in an inertial coordinate system.

(d) The limiting cases of fast internal relaxation and no internal relaxation are both considered for calculations of the angular momentum distribution and emissivity of spinning dust.

(e) Infrequent collisions of single ions which deposit an angular  momentum larger than the grain angular momentum prior to the collision are treated as Poisson-distributed events.

The wobbling disk-like grain has anisotropic rotational damping and excitation. Such an anisotropy can increase the peak emissivity by a factor $\sim 2$, and increases the peak frequency by a factor $1.4-1.8$, compared to the results from the DL98 model. 

Further improvements of the DL98 model were performed in Hoang et al.~\cite{2011ApJ...741...87H}, where a number of additional effects were taken into account:

(f) emission from very small grains of triaxial ellipsoid ({\it irregular}) shape with the principal moments of inertia $I_{1}\ge I_{2}\ge I_{3}$, (g) effects of the orientation of dipole moment $\bmu$ within the grain body for the different regimes of internal thermal fluctuations, (h) effects of compressible turbulence on the spinning dust emission.

The work found that a freely rotating irregular grain with a given angular momentum  radiates at multiple frequency modes. The resulting spinning dust spectrum has peak frequency and emissivity increasing with the degree of grain shape irregularity, which is defined by $I_{1}:I_{2}:I_{3}$. 

In the following, we discuss in more detail the most fundamental effects included in the spinning dust model.

\subsubsection{Internal relaxation within a torque-free motion}
In the absence of external torques, a spinning grain undergoes internal relaxation due to Barnett effect that converts its rotational energy into heat (vibrational energy). As a result, the axis of maximum moment of inertia $\hat{\ba}_{1}$ and the grain angular velocity $\bomega$ eventually become aligned with its angular momentum--a ground state of minimum rotational energy (Purcell 1979 \cite{1979ApJ...231..404P}). According to the Fluctuation-Dissipation theorem, there should exist a reverse process that transfers the grain vibrational energy into rotational energy, resulting in the fluctuations of the grain rotation from its ground state. For a grain of a few \AA, the Intramolecular Vibrational-Rotational Energy Transfer process (IVRET) occurs due to imperfect elasticity on a timescale $10^{-2}$ s,\cite{1979ApJ...231..404P}which is shorter than the IR emission time. So, when the vibrational energy changes due to photon absorption/IR emission, as long as the Vibrational-Rotational energy exchange exists, interactions between vibrational and rotational systems maintain a thermal equilibrium, i.e., $T_{\rm rot}\approx T_{\rm d}$ where $T_{\rm d}$ is vibrational dust temperature. As a result, the natural rotation configuration of the grain is along an intermediate axis not directed along $\hat{\ba}_{1}$.

The dynamics of a triaxial grain is more complicated than that of a axisymmetric grain. In addition to the precession of $\hat{\ba}_{1}$ around $\bJ$ as in the disk-like grain, $\hat{\ba}_{1}$ wobbles rapidly relative to $\bJ$, resulting in the variation of the angle between $\hat{\ba}_{1}$ and $\bJ$ (see Figure \ref{fig:3Dgrain}). To describe the torque-free motion of an irregular grain having a rotational energy $E_{\rm rot}$, the conserved quantities are taken, including the angular momentum $\bJ$, and a dimensionless parameter that characterizes the deviation of the grain rotational energy from its minimum value, $q={2I_{1}E_{\rm rot}}/{J^{2}}$. The value of $q=1$ corresponds to the ground state rotation around the axis of maximum moment of inertia.

\subsubsection{Power spectrum of a freely wobbling triaxial grain}

Consider a grain with a dipole moment $\bmu$ fixed in the grain body rotating with an angular momentum $\bJ$. If the grain only spins around its symmetry axis, then the rotating dipole moment emits radiation at a unique frequency $\nu$ equal to the rotational frequency, i.e., $\nu=\omega/2\pi$ (see DL98). The power spectrum for this case is simply a Delta function $\delta(\nu-\omega/2\pi)$
with a unique frequency mode.

\begin{figure*}
\centering
\includegraphics[width=0.8\textwidth]{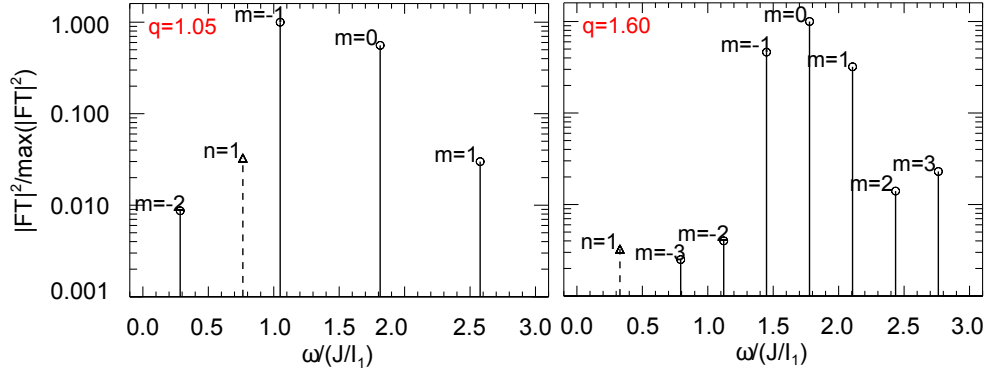}
\caption{\small Normalized power spectrum  of a torque-free rotating irregular grain with $I_{1}:I_{2}:I_{3}=1:0.6:0.5$ for the different values of $q=1.05$ and $1.60$. The components of $|{\rm FT}(\ddot{\mu}_x)|^2/\max(|{\rm FT}(\ddot{\mu}_x)|^2))$
(or $|{\rm FT}(\ddot{\mu}_y)|^2/\max(|{\rm FT}(\ddot{\mu}_x)|^2)$) are indicated by circles, while the components of $|{\rm FT}(\ddot{\mu}_z)|^2/ \max(|{\rm FT}(\ddot{\mu}_x)|^2)$ are indicated by triangles. Orders of in-plane modes $m$  and out-of-plane modes $n$ are indicated, and case 1 ($\mu_{1}=\mu/\sqrt{3}$) of $\bmu$ orientation is assumed. Figure reproduced from Hoang et al.\protect\cite{2011ApJ...741...87H}}
\label{fig:602}
\end{figure*}

For an irregular grain of triaxial ellipsoid shape, the grain rotational dynamics is more complicated. In general, one can also obtain analytical expressions for power spectrum, but it is rather tedious. To find the power spectrum of a freely rotating irregular grain, Hoang et al.~\cite{2011ApJ...741...87H} have employed a more simple, brute force approach based on the Fourier transform approach. The power spectra with numerous emission modes are shown in Figure \ref{fig:602} for two cases of small deviation ($q\sim 1$) and large deviation ($q=1.6$) from the ground state of minimal rotational energy. One can see that for the case with large $q\sim 1$, the emission power concentrates at the modes with $\omega/(J/I_{1})\sim 1$. For $q=1.6$, the emission power mostly distributes at modes with $\omega/(J/I_{1})\sim 1.8$. It indicates that if the grain rotational energy is increased so that the grain spends a significant fraction of time rotating with large $q$ (large deviation angle), then the grain will emit larger rotational emission.

\subsubsection{Distribution Function of Grain Angular Momentum}

To find the exact distribution function for grain angular momentum $\bJ$ subject to a variety of rotational damping and excitation processes, Hoang et al.~\cite{Hoang:2010jy,2011ApJ...741...87H} implemented a numerical approach based on the Langevin equation. Basically, they numerically solved the Langevin equations describing the evolution of three components of $\bJ$ in an inertial coordinate system, which are given by
\bea
dJ_{i}=A_{i}dt+\sqrt{B_{ii}}dq_{i},\mbox{~for~} i=\mbox{~x,~y,~z},\label{eq:612}
\ena
where $dq_{i}$ are random Gaussian variables with $\langle dq_{i}^{2}\rangle=dt$, 
$A_{i}=\langle {\Delta J_{i}}/{\Delta t}\rangle$  and $B_{ii}=\langle \left({\Delta J_{i}}\right)^{2}/{\Delta t}\rangle$ are damping and diffusion coefficients defined in the inertial coordinate system. The LE approach allows us to treat the spinning dust emission from wobbling grains with arbitrary temperature. Moreover, the impulsive excitation by single-ion collisions, which can deposit an amount of angular momentum greater than the grain angular momentum prior the collision, is easily included in Equation (\ref{eq:612}) using Monte-Carlo simulations.

\subsubsection{Emission from Wobbling Grains of Triaxial Ellipsoid Shape}
An irregular grain rotating with a given angular momentum $J$ radiates at frequency  modes $\omega_{k}\equiv \omega_{m}$ with $m=0,\pm 1, \pm 2...$ and $\omega_{k}\equiv \omega_{n}$ with $n=1,2,3...$\cite{2011ApJ...741...87H}. For simplicity, let denote the former as $\omega_{m_{i}}$ and the latter as $\omega_{n_{i}}$ where  $i$ indicates the value for $m$ and $n$. These frequency modes depend on the parameter $q(s)$, which is determined by the internal thermal fluctuations within the grain. 

To find the spinning dust emissivity by a grain at an observational frequency $\nu$, first we need to know how much emission that is contributed by each mode $\omega_{k}$. For a given angular momentum $J$, the probability  of finding the emission at the angular frequency $\omega$ is given by
\bea
pdf(\omega|J)d\omega=f_{\rm VRE}(s,J)ds= A{\exp}\left(-\frac{q(s)J^{2}}
{2I_{1}k_{\B}T_{\rm d}}\right)ds,~~~~~\label{eq:614}
\ena
which yields $pdf_{k}(\omega|J)=\left({\partial \omega_{k}}/{\partial s}\right)^{-1}
f_{\rm VRE}(s,J)$ for mode $\omega\equiv\omega_{k}(s)$. 

The emissivity from emission mode $k$ is calculated as
\bea
j_{\nu,k}^{a}=\frac{1}{4\pi}\int_{J_{l}}^{J_{u}}P_{\ed,k}(J,q_{\le})f_{J}(J)
pdf_{k}(\omega|J)2\pi~dJ+\frac{1}{4\pi}\int_{J_{l}}^{J_{u}}P_{\ed,k}(J,q_{>})f_{J}(J) 
pdf_{k}(\omega|J)2\pi~dJ,~~~~~\label{eq:615}
\ena
where $q_{\le}$ and $q_{>}$ denote $q\le q_{\sp}\equiv I_{1}/I_{2}$ and $q>q_{\sp}$, respectively; $J_{l}$ and $J_{u}$ are lower and upper limits for $J$ corresponding to a given angular frequency $\omega_{k}(J,q)=\omega$, and $2\pi$ appears due to the change of variable from $\nu$ to $\omega$.

The emissivity per H is obtained by integrating $j_{\nu}^{a}$ over the grain size distribution:
\bea
{j_\nu\over n_\H} = 
{1\over n_\H}
\int_{a_{\min}}^{a_{\max}} da {dn\over da}\sum_{k}j_{\nu,k}^{a}~,\label{eq:619}
\ena
where the summation is taken over all possible emission modes $\omega_{k}$.

Denote the parameter $\eta\equiv b_{3}/b_{2}=\alpha^{-2}$ where $b_{2}$ and $b_{3}$ are lengths of semiaxes $\hat{\ba}_{2}$ and $\hat{\ba}_{3}$ (see Figure \ref{fig:3Dgrain}), then the degree of grain shape irregularity is completely characterized by $\eta$. For each grain size $a$, the parameter $\eta$ is increased from $\eta=1$ to $\eta=\eta_{\max}$, where $\eta_{\max}$ is constrained such that the shortest axis $\hat{\ba}_{1}$ should not be shorter than the grain thickness $L$.

\begin{figure*}
\centering
\includegraphics[width=0.8\textwidth]{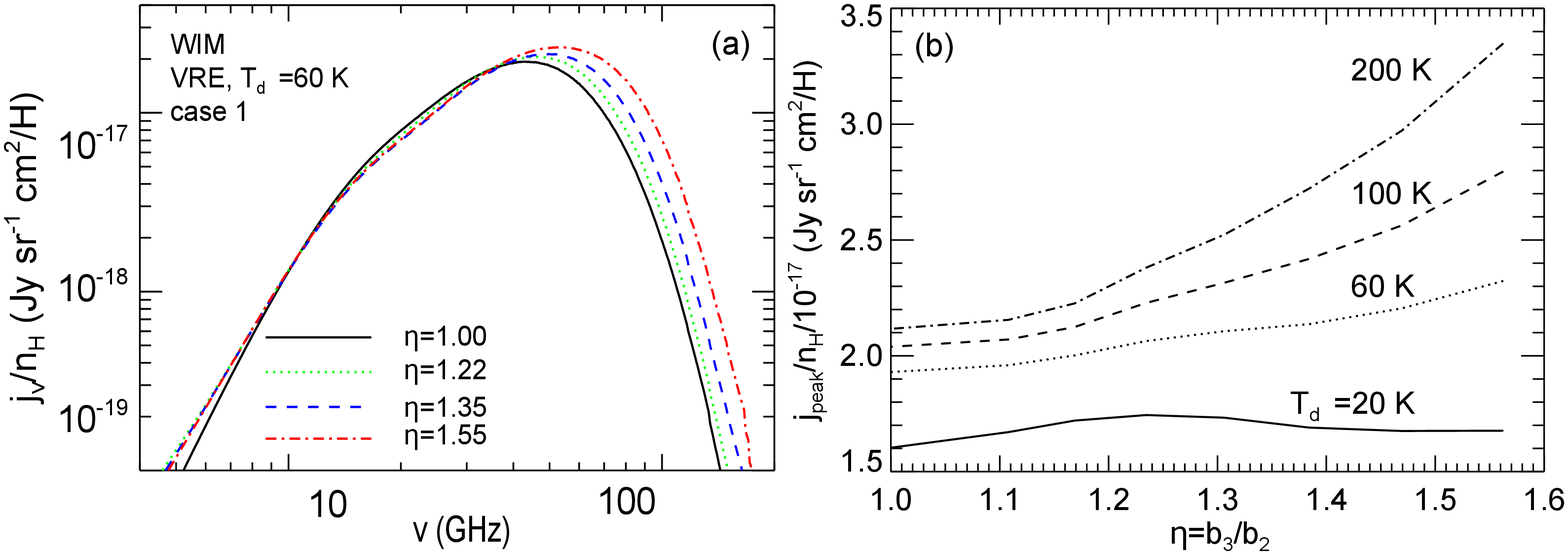}
\caption{\small (a) Emissivity per H from irregular grains of different degrees of irregularity $\eta=b_{3}/b_{2}$ with $T_{\rm d}=60$ K in the WIM for case 1 in which the electric dipole moment is isotropically oriented in the grain body. (b) Increase of spinning dust emissivity with the degree of grain irregularity for the different values of $T_{d}$. Figure reproduced from Hoang et al. \protect\cite{2011ApJ...741...87H}.}
\label{fig:604}
\end{figure*}

Figure \ref{fig:604}(a) shows the spinning dust emissivity for different degrees of irregularity $\eta$ and with a dust temperature $T_{\rm d}=60\K$ in the WIM. The emission spectrum for a given $T_{\rm d}$ shifts to higher frequency as $\eta$ decreases (i.e. the degree of  grain irregularity increases), but their spectral profiles remain similar. Figure  \ref{fig:604}(b) shows the increase of peak emissivity $J_{\rm peak}$ with increasing $\eta$. One particular feature in Figure \ref{fig:604}(b) is that for axisymmetric grains ($\eta=1$), the emissivity increases by a factor of $1.3$ with $T_{\rm d}$ increases from $20-200\K$. However, for the irregular grain with high triaxiality $\eta=1.5$, the emissivity increases by a factor of $2$. The peak frequency is increased by a factor of $1.4$. This arises from the fact that higher $T_{\rm d}$ can excite the grain to more emission modes, which allows more vibrational energy converted to rotational emission.

\subsubsection{Effect of transient single-ion collisions}
\begin{wrapfigure}{r}{0.5\textwidth}
\vspace*{-.1cm}
\includegraphics[width=0.5\textwidth]{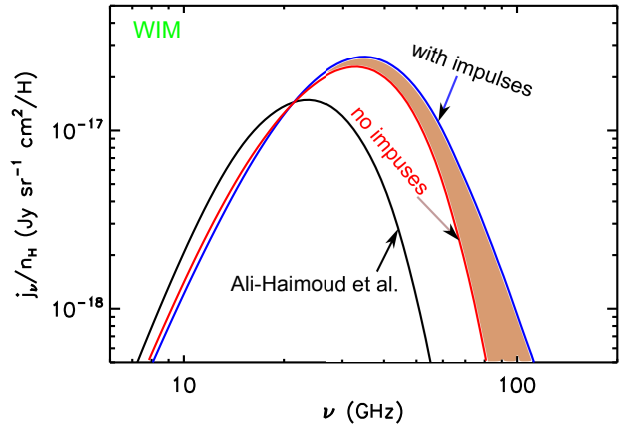}
\caption{\small Emissivity per H obtained for WIM without ionic impulses using the Fokker-Planck equation\protect\cite{2009MNRAS.395.1055A} and with impulses using our LE simulations for the wobbling grain. The spectra are efficiently broadened as a result of impulses (see blue line). Figure reproduced from Hoang et al.\protect\cite{Hoang:2010jy}.}
\label{f516}
\end{wrapfigure}

The effect of impulsive excitations by single-ion collisions is shown in Figure \ref{f516}. One can see that the impulses from ions can increase the emissivity by $\sim 23\%$, and slightly increase the peak frequency (see Figure \ref{f516}). The tail of high frequency part is obviously extended due to the contribution from ionic impulses with large angular momentum. 

\subsubsection{Spinning dust emission as a diagnostic tool of PAH physics}

Spinning dust emission involves a number of parameters, including grain physical parameters and environment parameters. Among them, the grain dipole moment is the most important parameters, which characterize the spinning dust emissivity \cite{2009ApJ...699.1374D,2011ApJ...741...87H}. Because the dipole moment of PAHs is dominated by the intrinsic dipole moment arising from polar bonds (e.g, C-H, C-CH$_{3}$ bonds), spinning dust emissivity may vary with local environments due to the dehydrogenation of PAHs by intense radiation near stars. 
\subsection{Other refined models of spinning dust emission}

Ali-Ha{\"\i}moud et al.~\cite{2009MNRAS.395.1055A} refined the original model of spinning dust and presented an analytic solution of the Fokker-Planck equation that describes the rotational excitation of a spherical grain. The effect of grain precession on rotational emission was independently studied in Silsbee et al.~\cite{Silsbee:2011p5567} using the Fokker-Planck equation approach. The aforementioned models disregarded the transient spin-up by infrequent single-ion collisions and considered two limiting cases of internal relaxation corresponding to $T_{d}\rightarrow 0$ and $T_{d} \rightarrow \infty$. Ysard et al.~\cite{2010A&A...509A..12Y} revisited the spinning dust emission using the quantum mechanic approach, which is in good agreement with the classical approach in the DL98 model.

\section{Polarization of Spinning Dust Emission}
\subsection{Alignment of PAHs by resonance paramagnetic relaxation}
Quantitative predictions for the polarization of spinning dust emission is particularly useful for analysis of CMB polarization data. It is noted that the electric field of spinning radiation lies in the plane perpendicular to the rotational axis, which varies constantly due to the randomization by gas collisions and IR emission. As a result, the net polarization is determined by the degree of alignment of PAHs with the magnetic field. The first study exploring the paramagnetic alignment of PAHs was performed by Lazarian and Draine \cite{2000ApJ...536L..15L}. They identified a new process of relaxation, which was termed {\it resonance paramagnetic relaxation} and which is orders of magnitude more efficient for the rapidly rotating PAHs compared to the classical Davis-Greenstein process. Hoang et al. \cite{Hoang:2014cw} have computed the degree of alignment of PAHs due to both paramagnetic relaxation and resonance relaxation. Further calculations are conducted in Hoang and Lazarian~\cite{Hoang:2015ts} for the different magnetic field strengths. The predicted polarization is below $5\%$ for the typical conditions of the ISM.

\subsection{Constraints from Ultraviolet Polarization of Starlight}
\begin{wrapfigure}{r}{0.5\textwidth}
\vspace*{-1.cm}
\includegraphics[width=0.45\textwidth]{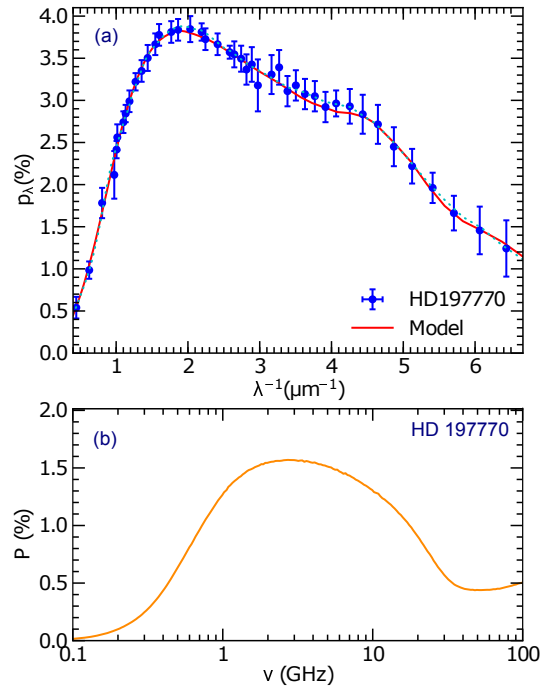}
\caption{\small (a) Best-fit model to the polarization curve of starlight toward HD 197770 with the UV polarization bump. (b) Polarization of spinning dust emission obtained with the degree of alignment of PAHs from the best-fit model. Figure reproduced from Hoang et al.\protect\cite{2013ApJ...779..152H}.}
\label{fig:f5}
\end{wrapfigure}

Since the PAHs that emit the spinning dust emission are likely to be the same carriers that produce the UV extinction bump at 217.5 nm (see, e.g., Draine and Li \cite{2007ApJ...657..810D}), one can look for the signature of aligned PAHs using the starlight polarization at this wavelength. The polarization bump at 217.5 nm toward two stars, HD 197770 and HD 147933-4, was discovered a long time ago \cite{1992ApJ...385L..53C,1993ApJ...403..722W}. Fitting to the observational data for these stars, Hoang et al. \cite{2013ApJ...779..152H} found that a model with aligned silicate grains plus weakly aligned PAHs can successfully reproduce the UV polarization bump as well as the polarization plateau. With the weak alignment inferred, spinning dust emission is found to be weakly polarized, with the upper limit of polarization of $\sim 1.6\%$ (see Figure \ref{fig:f5}). The derived constraint is consistent with the observational data \cite{2011MNRAS.418L..35D,2011MNRAS.418..888M,LopezCaraballo:2011p508,RubinoMartin:2012ji}. Latest results by Planck~\cite{2015arXiv150606660P} also show the upper limits of AME polarization to be $0.6\pm 0.5$ percent level. 

\section{Alternative sources of AME}
Numerous observations have indicated that AME is most likely produced by rapidly spinning tiny dust grains (see PlanckCollaboration \cite{2015arXiv150606660P}).  Although it is usually attributed to spinning PAHs, the precise carrier of spinning dust emission is still unclear. For instance, the analysis of WISE all-sky data in Hensley et al. \cite{Hensley:2015wc} found uncorrelation of AME with PAH abundance, which is interpreted as a case against spinning PAHs. Moreover, Bernstein et al. \cite{Bernstein:2015ct} show the correlation of AME with diffuse interstellar bands (DIBs), suggesting that AME is likely produced by the same carriers as DIBs. More observational, modeling and theoretical studies should be done to understand better the origins of AME.

In addition to spinning dust emission,  magnetic dipole emission (hereafter MDE) from magnetic particles was also suggested as a possible source of AME by Draine and Lazarian \cite{1999ApJ...512..740D}. Recently, Draine and Hensley \cite{2013ApJ...765..159D} refined the MDE model by using the Gilbert equation approach and find that MDE is an important foreground for frequencies $\nu=70-300$ GHz and contributes little to AME unless magnetic nanoparticles are extremely elongated. It is noted that the MDE by free-flying magnetic particles was frequently ruled out as a source of AME appealing to its high expected polarization level \cite{2011MNRAS.418L..35D,LopezCaraballo:2011p508,GenovaSantos:2015hc}. Recent results in Hoang and Lazarian \cite{Hoang:2015ts} show that the polarization of MDE from free-fliers is rather low, consistent with the observed polarization of AME. Thus, the magnetic origin by free-flying nanoparticles may be more important that previously thought.

\section{Summary}

The principal points of this paper are summarized as follows:

(1) The model of spinning dust emission proposed by DL98 proved capable of explaining anomalous microwave emission, and its predictions were supported by numerous observations since the introduction of the model.

(2) The DL98 spinning dust model has been improved by including the effects of thermal fluctuations within dust grains, impulsive excitations with single ions, transient heating by UV photons, triaxiality of grain shape, and compressible turbulence, which made the spinning dust model more realistic.

(3) Spinning dust emission involves a number of grain physical parameters (e.g., dipole moment) and environment parameters (PAH abundance, gas density). With the latest progress on theoretical modeling and observations, the possibility of using spinning dust as a diagnostics tool for physical parameters of ultrasmall dust is open.

(4) Our numerical calculations and modeling show that spinning dust emission from PAHs is weakly polarized, which is consistent with the available observational data of the AME polarization.

\section*{Acknowledgments}
I am very grateful to Alex Lazarian, Bruce Draine, and Peter Martin for fruitful collaborations on this interesting topic. T.H. acknowledges the support by Alexander von Humboldt Foundation as a Postdoctoral Fellowship at Ruhr Universit$\ddot{\rm a}$t Bochum and Goethe Universit$\ddot{\rm a}$t Frankfurt am Main, Germany.

\section*{References}
\small
\bibliography{ms.bbl}

\begin{thebibliography}{10}

\bibitem{Kogut:1996p5303}
A~Kogut, et al.,
\newblock {\em \apj}, 460,1,1996.

\bibitem{1997ApJ...482L..17D}
A de~Oliveira-Costa, A~Kogut, M~J Devlin, C~B Netterfield,
  L~A Page, and E~J Wollack,
\newblock {\em \apjl}, 482, L17, 1997.

\bibitem{Leitch:1997p7359}
E~M Leitch, A~C~S Readhead, T~J Pearson, and S~T Myers,
\newblock {\em \apjl}, 486, L23, 1997.

\bibitem{deOliveiraCosta:1998p4707}
A de~Oliveira-Costa, M Tegmark, L~A Page, and S~P
  Boughn,
\newblock {\em \apj}, 509:L9, 1998.

\bibitem{deOliveiraCosta:1999p4706}
A de~Oliveira-Costa, M Tegmark, C~M Gutierrez, and et al,
\newblock {\em \apj}, 527:L9, 1999.

\bibitem{1999ASPC..181..253M}
P~R McCullough, J~E Gaustad, W~Rosing, and D~Van~Buren,
\newblock {\em ASPC}, 181:253, 1999.

\bibitem{1998ApJ...508..157D}
B~T Draine and A~Lazarian,
\newblock {\em \apj}, 508:157, 1998.

\bibitem{deOliveiraCosta:2000p4667}
A de~Oliveira-Costa, Max Tegmark, Mark~J Devlin, and et al,
\newblock {\em \apj}, 542:L5, 2000.

\bibitem{2002ApJ...566..898F}
D~P Finkbeiner, D~J Schlegel, C Frank, and C Heiles,
\newblock {\em \apj}, 566:898, 2002.

\bibitem{Gold:2009p5268}
B~Gold, C~L Bennett, R~S Hill, and et al,
\newblock {\em \apjs}, 180:265, 2009.

\bibitem{Gold:2011p5261}
B~Gold, N~Odegard, J~L Weiland, and et al.,
\newblock {\em \apjs}, 192:15, 2011.

\bibitem{PlanckCollaboration:2011p515}
{Planck Collaboration}, P~A~R Ade, and et al.,
\newblock {\em A\&A}, 536:A20, 2011.

\bibitem{Scaife:2010p569}
{AMI Consortium}, A M~M Scaife, B Nikolic, D~A Green, and et~al.,
\newblock {\em \mnras}, 406::45, 2010.
    
\bibitem{2012ApJ...754...94T}
C~T Tibbs, R~Paladini, M~Compi{\`e}gne, and et al.,
\newblock {\em \apj}, 754:94, 2012.

\bibitem{2008MNRAS.391.1075C}
S Casassus, C Dickinson, K Cleary, and et al.,
\newblock {\em \mnras}, 391:1075, 2008.

\bibitem{2011MNRAS.418.1889T}
C~T Tibbs, N~Flagey, R~Paladini, and et al.,
\newblock {\em \mnras}, 418:1889, 2011.

\bibitem{2010A&A...523A..20B}
C~Bot, N~Ysard, D~Paradis, and et al.,
\newblock {\em A\&A}, 523:20, 2010.

\bibitem{Hoang:2010jy}
Thiem Hoang, B~T Draine, and A~Lazarian,
\newblock {\em \apj}, 715:1462, 2010.

\bibitem{2011ApJ...741...87H}
Thiem Hoang, A~Lazarian, and B~T Draine,
\newblock {\em \apj}, 741:87, 2011.

\bibitem{1979ApJ...231..404P}
E~M Purcell.
\newblock {\em \apj}, 231:404, 1979.

\bibitem{2009MNRAS.395.1055A}
Y Ali-Ha{\"\i}moud, C~M Hirata, and C Dickinson,
\newblock {\em \mnras}, 395:1055, 2009.

\bibitem{2009ApJ...699.1374D}
G Dobler, B Draine, and D~P Finkbeiner,
\newblock {\em \apj}, 699:1374, 2009.

\bibitem{Silsbee:2011p5567}
K Silsbee, Y Ali-Ha{\"\i}moud, and C~M Hirata,
\newblock {\em \mnras}, 411:2750, 2011.

\bibitem{2010A&A...509A..12Y}
N~Ysard and L~Verstraete.
\newblock {\em A\&A}, 509:12, 2010.

\bibitem{2000ApJ...536L..15L}
A~Lazarian and B~T Draine.
\newblock {\em \apj}, 536:L15, 2000.

\bibitem{Hoang:2014cw}
Thiem Hoang, A~Lazarian, and P~G Martin,
\newblock {\em \apj}, 790:6, 2014.

\bibitem{Hoang:2015ts}
Thiem Hoang and A~Lazarian,
\newblock {\em arxiv:1511.03691}, 2015.

\bibitem{2007ApJ...657..810D}
B~T Draine and Aigen Li.
\newblock {\em \apj}, 657:810, 2007.

\bibitem{1992ApJ...385L..53C}
G~C Clayton, C~M Anderson, A~Mario Magalhaes, and et al,
\newblock {\em \apj}, 385:L53, 1992.

\bibitem{1993ApJ...403..722W}
M~J Wolff, G~C Clayton, and M~R Meade,
\newblock {\em \apj}, 403:722, 1993.

\bibitem{2013ApJ...779..152H}
Thiem Hoang, A~Lazarian, and P~G Martin,
\newblock {\em \apj}, 779:152, 2013.

\bibitem{2011MNRAS.418L..35D}
C~Dickinson, M~Peel, and M~Vidal,
\newblock {\em \mnras}, 418:L35, 2011.

\bibitem{2011MNRAS.418..888M}
N~Macellari, E~Pierpaoli, C~Dickinson, and J~E Vaillancourt,
\newblock {\em \mnras}, 418:888, 2011.

\bibitem{LopezCaraballo:2011p508}
C~H L{\'o}pez-Caraballo, J~A Rubi{\~n}o-Mart{\'\i}n, R~Rebolo, and
  R~G{\'e}nova-Santos,
\newblock {\em \apj}, 729:25, 2011.

\bibitem{RubinoMartin:2012ji}
J~A Rubi{\~n}o-Mart{\'\i}n, C~H L{\'o}pez-Caraballo, R~G{\'e}nova-Santos, and
  R~Rebolo,
\newblock {\em Advances in Astronomy}, 2012:1, 2012.

\bibitem{2015arXiv150606660P}
Planck Collaboration, P~A~R Ade, N~Aghanim, and et al.,
\newblock {\em arXiv:1506.06660}, page 6660, 2015.

\bibitem{Hensley:2015wc}
Brandon~S Hensley and B~T Draine,
\newblock {\em arXiv.org}, 2015.

\bibitem{Bernstein:2015ct}
L~S Bernstein, F~O Clark, J~A Cline, and D~K Lynch,
\newblock {\em \apj}, 813:1, 2015.

\bibitem{1999ApJ...512..740D}
B~T Draine and A~Lazarian,
\newblock {\em \apj}, 512:740, 1999.

\bibitem{2013ApJ...765..159D}
B~T Draine and Brandon Hensley,
\newblock {\em \apj}, 765:159, 2013.

\bibitem{GenovaSantos:2015hc}
R~G{\'e}nova-Santos, J~A~Rubi{\~n}o Mart{\'\i}n, R~Rebolo, and et~al.,
\newblock {\em \mnras}, 452:4169, 2015.

\end{thebibliography}


\end{document}